\documentstyle[12pt,aasms4]{article}
\newcommand{\kms}{km~s$^{-1}$}
\newcommand{\al}{$\alpha$}
\newcommand{\lam}{$\lambda$}
\newcommand{\cm}{cm$^{-2}$}

\begin{document}

\title{Molecular Hydrogen Absorption in the z = 1.97 Damped Ly\al\ 
Absorption System toward QSO 0013$-$004\footnote{Observations here were obtained with the
Multiple Mirror Telescope, a joint facility of the University of Arizona and the Smithsonian Institution.}}

\author{Jian Ge \& Jill Bechtold}
\affil{Steward Observatory, University of Arizona, Tucson, AZ 85721}
\authoremail{jge@as.arizona.edu; jill@as.arizona.edu}

\begin{abstract}
We present  a new ultra-violet spectrum of the QSO 0013$-$004 with 0.9 \AA\ 
resolution obtained with the MMT Blue  spectrograph. The $\upsilon$ = 0 - 0,
1  - 0, 2 - 0 and 3 - 0 Lyman  bands of H$_2$  associated with
the z = 1.9731 damped Ly\al\ absorption line system have been detected.  The
  H$_2$ column density is N(H$_2$) = 6.9$(\pm 1.6)\times 10^{19}$ \cm, 
and the 
Doppler parameter b = 15$\pm 2$ \kms. The
populations of different rotational levels are measured and 
used to derive the excitation temperatures.
 The estimated kinetic temperature 
T$_K\sim 70$ K,
 and the total
particle number density n$(H) \sim$ 300 cm$^{-3}$. The UV photoabsorption
rate $\beta_0 \sim 6.7\times 10^{-9}$ s$^{-1}$, about a factor of few times
 greater than
that in a typical diffuse Milky Way interstellar cloud.   The 
total  hydrogen column density 
is N(H) = 6.4($\pm 0.5)\times 10^{20}$ \cm. The fractional H$_2$ abundance
$f$ = 2N(H$_2$)/(2N(H$_2$) +  N(H I)) $\sim$ 0.22 $\pm$ 0.05
 is the highest among all 
observed damped Ly\al\ absorbers. The high fractional H$_2$ abundance
 is consistent with the inferred presence of dust and strong C I absorption
in this absorber.  
\end{abstract}

\keywords{ISM: molecules -- quasars: absorption lines -- 
quasars: individual (Q0013$-$004)}

\section{Introduction}

Damped Ly$\alpha$ quasar absorption line systems are commonly believed to 
trace a population of objects which may be  the 
progenitors of modern  galaxies  (e.g. 
Wolfe 1990).  
The typical  metallicity and dust-to-gas ratio in the damped Ly$\alpha$ 
systems at z $\sim$ 2 is measured to be 
about 10\% of those  in the Milky Way with a large scatter 
(e.g. Pettini {\em et al.}   1994; 1995;
 Pei, Fall \& Bechtold, 1991; Lu {\em et al.}   1996).
The UV  Lyman and Werner bands of H$_2$ have been difficult to detect 
however,  so that little
is known about molecular gas in damped Ly\al\ galaxies
(Levshakov {\em et al.}   1992; Lanzetta {\em et al.}   1989; Foltz et al
1988; Chaffee {\em et al.}   1988;  Black {\em et al.}   1987).    The problem such searches face  is that without high resolution and high 
signal-to-noise ratio
spectra, the molecular hydrogen lines are  blended with the Ly\al\
forest lines. Further, quasar absorption line studies usually concentrate
on bright blue quasars, so lines of sight with substantial reddening and large molecular
fraction are probably selected against
(Ostriker \& Heisler, 1984; Fall \& Pei 1993).  So far, there is only one detection of H$_2$ absorption 
 in the z = 2.811 damped Ly\al\ absorber toward PSK 0528$-$250,
which  shows moderate H$_2$ absorption with total column density  
N(H$_2$) = $1\times 10^{18}$ 
\cm\ (Foltz {\em et al.}   1988; Lanzetta 1993). This system
may not be representative since its redshift is very close to the redshift
of the quasar.

Previous observations of the z$_{ab}$ = 1.9731 damped Ly\al\ absorber
toward Q 0013$-$004 (z$_{em}$ = 2.0835) showed C I absorption, and line ratios  of Cr II, Fe II and Zn II
indicated significant depletion, and hence a relatively high 
dust-to-gas ratio compared to other damped Ly\al\ absorbers (Pettini et al.
1994, Ge, Bechtold, \& Black 1996). Thus, it seemed to be a good system to 
search for H$_2$. 
In this {\it Letter}, we  report the detection of
strong  H$_2$ absorption lines.

\section{Observations}
Spectra were obtained with the Multiple Mirror Telescope (MMT)
on Oct. 13  and Oct. 24 - 25, 1995 with the Blue Channel Spectrograph and the 
Loral 3072$\times$1024 CCD.  The 832 l/mm  grating was used  in
second order.  A 1$''\times 180''$ slit was used   to give a 
resolution of 0.9 \AA, corresponding to  2.6  pixels, measured from the comparison lamp lines. The wavelength
coverage was from 3000 \AA\ to 4100 \AA.  The 
seeing was 1.0$''$ (FWHM). The integration  was composed of twelve separate
exposures  of 50 minutes each. The QSO was stepped by a few arcseconds  along the slit  between each
exposure   to smooth out any residual irregularities in the detector
response which remained after flat-fielding. An exposure  of a He-Ne-Ar-Cu
lamp  was done before and after each exposure of the object
to  provide  an accurate wavelength reference, and a measure  of the spectral
resolution.  Several quartz lamp exposures were taken at the beginning and end of the nights to provide  a flat-field correction.
 We  also observed
standard stars before and after the QSO observations to monitor  possible 
atmospheric absorption.

 The data
were reduced in the standard  way using IRAF.
  The He-Ne-Ar-Cu spectra
were extracted with the  same procedure and a first order Spline3 function
fit was used to obtain a wavelength calibration.  The root-mean-square
 residuals to the
wavelength fit are typically 0.05 \AA.
 All reported wavelengths 
are vacuum and have been corrected to the heliocentric frame.
We have rebined pixels in the 
common wavelength  ranges of the spectra from the  two observation  runs
and  combined them, weighted by the signal-to-noise ratio (S/N). Part of the summed spectra are shown
in Figures 1(a) and 2.
 The  S/N is  $\sim$ 10 per pixel in   the wavelength range of 3150 
 - 3350 \AA, which covers the H$_2$ absorption bands.  The S/N $\sim$ 20 or 
higher in  the  wavelength range of 
3500 - 3750 \AA, which covers the damped Ly\al\ absorption.
  The continuum was fitted in the way 
described by Bechtold (1994). The spectra shown were normalized by their fitted 
continuum.

\section{Results}

Fig. 1(a)   shows the spectral region containing the H$_2$ Lyman $\upsilon$ = 0 - 0,
 1 - 0, 2 - 0 and 3 - 0 bands for the z = 1.9731 damped Ly\al\ absorber. A 
simulated spectrum of the H$_2$ Lyman bands is plotted as a dotted line for
comparison. There is  
remarkable  agreement between  the two spectra  despite the presence of Ly\al\ 
forest lines.

 In order to investigate whether   
 the observed configuration
of lines is the result of chance  coincidence with   Ly\al\ forest lines,
  we cross-correlated the observed spectrum (3150 - 3350 \AA)
 with  the simulated H$_2$ absorption line spectrum  
(3150 - 3350 \AA)
using FXCOR in IRAF (see Tonry \& Davis 1979 section III for details;
 Foltz et al., 1988).
The cross-correlation  function was computed for lags in the range 
$-$16,000 $\le \Delta v \le 16,000$ \kms\ and is shown in Fig. 1(b).
 A very strong peak with the correlation amplitude of 0.66 is  noted 
 near zero velocity  shift. The other two 
second strong peaks at $\sim \pm$ 4,300 \kms\ are caused by shifting the 
spectra by one band separation ($\sim$ 47 \AA\ in the observed frame, 
e.g. Morton \& Dinerstein,  1976). If the regions in the spectrum around the
 four identified strong  
Ly\al\ and metal lines, as marked in Fig. 1(a) (3216.2 - 3224.9 \AA, 3296.1 - 
3300.3 \AA),  are removed from the data 
before the correlation was carried out, the peak correlation amplitude is 0.82.
The measured redshift for the H$_2$
absorption corresponding to the peak position of  the cross-correlation function,
 z$_{H_2}$ = 1.9731, is consistent with the redshift of the  metal
lines (Ge {\em  et al.} 1997).

In order  to estimate  the statistical significance of the correlation
amplitude,  we have simulated  Ly\al\ forest line spectra 
 using the program described  in Dobrzycki \& Bechtold 
(1996). In order to be conservative, we made  synthetic spectra
that matched the  total absorption line density of all the strong lines  in the
Q 0013$-$004 spectrum  between 3150 and 3350 \AA\  even though several of 
the strong lines are actually metal transitions for the z = 1.97 system.
 We  cross-correlated  1000 simulated  spectra with the synthetic 
H$_2$ absorption line spectrum. The
 peak values of the 
cross-correlation function are all smaller than 0.55. Thus, the 
probability of obtaining a correlation peak of 0.66 by chance is less than
10$^{-4}$.

To obtain column densities from the measured values of equivalent width (Table 
1), 
a curve-of-growth was constructed for J = 2, 3, 4, 5 rotational levels 
(Fig. 3). 
The theoretical  curves for the 0-0 Lyman band with different Doppler b-values,
b = 10, 13, 15,  17, 20 \kms, are shown.  Absorption lines for J =  4, 5 are on the
linear section. Absorption lines from J = 2, 3 are on the flat section.  The observed values for J = 2, 3, 4 and 5 are consistent with b = 15$\pm$2 
\kms. 
 Table 2 shows the derived column densities for J =  2, 3, 4
assuming the b = 15$\pm$2 \kms, and upper  limits for J = 5, 6, 7 assuming
they are on the linear portion of the curve-of-growth.
 Column densities from J = 0, 1 are derived by
the continuum-reconstruction procedure  described by Savage 
{\em et al.}    (1977), i.e. a reconstruction of the QSO spectrum 
  was  generated by 
dividing the observed spectrum by the synthetic H$_2$ spectrum of R(0), R(1), 
and P(1) lines. The final values of N(J=0) and N(J=1) were determined from
the best-fit reconstructions of  (2,0), (1,0) and (0,0) Lyman bands (Table 2).
The uncertainties in the final column densities were estimated by the same 
procedure. Together, the total H$_2$
 column density is N(H$_2$) = 6.9$(\pm 1.6) \times 10^{19}$ \cm.

Table 2 also includes the derived excitation temperatures for 
the rotational levels. 
 The measured 
T$_{01} = 70\pm 13$ K is in the range of typical values of the
 Milky Way diffuse clouds (Savage {\em et al.}   1977) and also similar  to the 
z = 2.811 damped Ly\al\ absorber of PKS 0528$-$250 (Songaila \& Cowie  1995).
 The  T$_{02} = 82^{+17}_{-6}$ K is  also in the range of   the
average value for the Milky Way clouds. The  excitation temperatures for 
higher J, T$_{ex}$, of 200 - 500 K, are similar to that of
 Milky  Way clouds (Spitzer {\em et al.}   1974).

Fig. 2 shows the observed damped Ly\al\ profile and  fits to the line profile.
 The  neutral hydrogen column density   N(H I) =  5.0($\pm  0.5)\times 
10^{20}$ \cm,  the Doppler parameter, b$_{H I}$ = 50 \kms\ and
 z$_{H I}$ = 1.9731. The neutral hydrogen absorption redshift
 is consistent   with the metal  line and H$_2$ 
redshifts.  The measured
fractional H$_2$ abundance $f$ = 2N(H$_2$)/(2N(H$_2$) +  N(H I)) =
0.22 $\pm$ 0.05,  is about a  factor of 60 times higher than that of  the 
z = 2.811 damped Ly\al\ system  toward PKS 0528$-$250 and the limits for other 
damped systems previously searched for H$_2$ absorption (Black et al,
1987; Chaffee {\em et al.}    1988, Foltz {\em et al.}    1988; Lanzetta {\em et al.}    1989; Levshakov
{\em et al.}    1992; Songaila \& Cowie, 1996).

\section{Discussion}

The molecular fraction $f$ = 0.22 $\pm$ 0.05  for the z = 1.97 absorber 
 is similar to 
that seen in E(B  $-$ V) $>$ 0.1 clouds in the Milky Way (Savage 
{\em et al.}   1977).  Indeed, the upper limit of the relative 
depletion of Cr to Zn,  [Cr/Zn]   $\le$ $-$ 1.0 (Pettini {\em et al.}   1994) 
suggests that the dust-to-gas ratio in this absorber is much 
higher than the average value of 10\% of the Milky Way's ratio 
for damped systems at z $\sim$ 2 (Pettini
 {\em et al.} 1994).
 The previous observation of Fe II \lam\  1608.45 \AA\  provides a direct comparison of the
  relative  depletion of Fe to Zn, [Fe/Zn] = $-$ 1.2 (Ge {\em et al.}   1997) to those
of the Milky Way's diffuse clouds. Table 3 shows a
comparison  between the z = 1.9731 absorber and 
the diffuse clouds toward $\zeta$ Oph and $\xi$ Per (Spitzer {\em et 
al}.  1974; Savage {\em et al.} 1977; Jura 1975; Savage {\em et al} 1992; 
Cardelli {\em et al.} 1991). 
 The relative depletion of [Fe/Zn] implies that the dust-to-gas
ratio in  the z = 1.9731 absorber is 
 $\sim$ 50\% of the Milky Way's value. The relatively normal dust-to-gas ratio
is consistent with the  high H$_2$ fraction because H$_2$ is very
 efficiently formed 
 on  dust grain surfaces (Savage {\em et al.}   1977). 

As discussed by  previous reviews  (e.g. Spitzer \& Jenkins, 1975; 
Shull \& Beckwith 1982), the relative populations  of J = 0 and 1 are 
established dominantly by thermal particle collisions, especially for 
saturated lines, so the excitation temperature, T$_{01}$, is approximately
equal to the kinetic temperature, T$_k$, of the clouds. The measured 
T$_{01} = 70\pm 13$ K in the z = 1.9731 absorber implies that the kinetic temperature
is similar to that of the Milky Way diffuse clouds, $\langle T_{01}\rangle$
= (77 $\pm$ 17) K (Savage {\em et al.}   1977). 
The  higher rotational levels are  populated primarily by collisions, 
formation pumping and UV pumping and radiative cascade after photoabsorption to the Lyman
and Werner bands (e.g. Spitzer \& Zweibel 1974; Jura 1974, 1975).  For example,
the J = 4 is populated by direct formation pumping, and by UV pumping from J = 0
(e.g. Black \& Dalgarno 1976). For densities less than $10^4$ cm$^{-3}$, the 
J = 4 level is depopulated  mainly by  spontaneous emission (Dalgarno \& Wright
1972; Elitzur \& Watson 1978).  Therefore, in a steady state for J =  4,
\begin{equation}
p_{4,0}\beta(0)n(H_2,J = 0) + 0.19 Rn(H)n = A_{42}n(H_2, J = 4),
\end{equation}
where  $p_{4,0}  = 0.26$ is the UV pumping efficiency into the J = 4 level from
the J = 0 level (Jura 1975),  $A_{42} = 2.8\times 10^{-9}$ s$^{-1}$ is the
 spontaneous transition probability (Dalgarno \& Wright 1972), R is the H$_2$
formation rate, n = n(H) + 2n(H$_2$), and $\beta(J)$ denotes the rate of  
absorption in the Lyman and Werner  bands from the Jth rotational level,
 including any attenuation (see Jura 1975 for details).  
 The equilibrium between  
the H$_2$ formation   on dust grains and H$_2$ destruction by absorption of Lyman- and Werner-band
radiation (e.g. Jura  1975) can be written as
\begin{equation}
I n(H_2) = R n(H)n \approx 0.11 \sum_{J=0}^6 \beta(J) n(H_2,J),
\end{equation}
where $I$ is the H$_2$ dissociation rate (Jura 1975). If  the self-shielding
in J = 0, 1 levels  is about the same, so that $\beta(0)\approx \beta(1)$, then 
\begin{equation}
n(H_2,J=4)A_{42} = 1.52 Rn(H)n.
\end{equation}
Thus, $Rn = 8.1(^{+3.9}_{-2.6})\times 10^{-15}$ s$^{-1}$ for the Q 0013$-$004 cloud, which is about the same magnitude as that for $\xi$ Per and
$\zeta$ Oph clouds
 (Jura 1975).

We can  use the analytic calculation for n(H$_2$)/n(H)  within a H$_2$
 cloud by Jura (1974) to estimate the H$_2$ 
dissociation rate $I \approx  7.4\times 10^{-10}$ s$^{-1}$.
The photoabsorption rate  in the Lyman  and Werner bands outside of the
cloud, $\beta_0 \approx I/0.11 \approx 6.7 \times 10^{-9}$ s$^{-1}$,
which is  
similar to that of  $\xi$ Per cloud (Jura 1975) and  about a  factor of a few
higher than that of $\zeta$ Oph cloud (Federman {\em et al.} 1995). Further,
the 
photoabsorption rate, $\beta_0$,   depends linearly on the local radiation field  at 930 
- 1150 \AA\ (Jura 1974),  therefore, the $\beta_0$ for the z =  1.9731 absorber
corresponds to an estimated local 
radiation field at 1000 \AA\  of J$_{1000 \AA}$
 $\approx 3\times 10^{-18}$
ergs cm$^{-2}$s$^{-1}$Hz$^{-1}$ster$^{-1}$. Thus J$_{1000\AA}$ is about three
 orders of 
magnitude higher than the radiation field at the Lyman limit, J$_{912\AA}$, expected  in the ambient   IGM
 at this  redshift  (e.g. Lu {\em et  al.} 1991; Bechtold 1994). The 
radiation  field at 1000 \AA\ is
 therefore probably  dominated by the UV emission by 
hot stars in this galaxy.

As mentioned above, the z = 1.9731 absorber has a similar 
dust-to-gas ratio to that of  the Milky Way diffuse clouds.   If we assume
the  H$_2$ formation  rate  on grains,  R $\sim 3\times 10^{-17}$ 
cm$^{3}$s$^{-1}$, the typical rate for the Milky Way's clouds
(Jura 1975), then the inferred value of the number density 
 n $\sim$ 300  cm$^{-3}$,
about the same  as  that of  the $\zeta$ Oph and $\xi$ Per clouds.

The derived values of the  UV radiation field, density, and  temperature in 
the z = 1.97 absorber are estimates based  on the simple analysis  of Jura  
(1975), which ignores  the depth-dependence  of the attenuation by dust
and self-shielding of absorption  lines. 
 However, the results from  this simple analysis
 are qualitatively consistent with that from  more detailed modeling (e.g.
van Dishoeck  \& Black 1986).

We note that  there is some
uncertainty in the derived b-value. While b = 15 \kms\ provides the best 
fit to the
observed values, smaller Doppler parameters, for  example,  b = 2 \kms,   also give an acceptable solution. In this case, the implied  
column densities in J = 2, 3 and  4 are so  high  they would suggest that
conditions more like a photon-dominated region (PDR)  are  implied 
(e.g. Draine \& Bertoldi 1996; Black \& van Dishoeck 1987;
Abgrall {\em  et al.} 1992; Le Bourlet {\em  et al.} 1993;  Sternberg \& 
Dalgarno 1989). The populations in J = 2, 3 and 4 
would be fit by a single excitation temperature  of 350 K.
  This would leave  excess population in J =  0 and 1, suggesting
 a cold component at  T$_{01}\approx 63$ K. The density of the
absorber cloud would have to
be lower than $\sim$ 5000 cm$^{-3}$ in order for the J = 5 limit to be 
consistent with the populations in J = 2, 3 and 4. 

Finally, if the  larger value b = 15 \kms\ is correct, then this  absorber has
 a b-value which is much larger than that for  typical Milky  Way
diffuse  clouds (e.g. Spitzer {\em et al.} 1974).  This would  indicate
 that there is likely
 more  than  one velocity component.  
 Spectra  of higher resolution would
permit a  better constrained analysis  of the  excitation  and molecular  
abundance and provide a better understanding of physical conditions 
in this high redshift galaxy.


\acknowledgements
We thank Dr. John Black for providing his H$_2$ absorption line synthesis code
and helpful discussions. We thank Dr. Adam Dobrzycki for use of his spectral
synthesis code. We thank Dr. Dave Meyer for helpful conversations.
 We  thank the staff of MMTO for their  help. 
This research was supported by NSF AST-9058510.

\newpage
\rm
\scriptsize
\begin{flushleft}
Table 1. H$_2$ absorption lines in the z = 1.9731 damped Ly$\alpha$ absorber  
\end{flushleft}
\begin{tabular}{llllll}  \hline\hline
No. & $\lambda_{rest}$(\AA) & $\lambda_{obs}$(\AA) & W$_{obs}$(\AA)&z$_{abs}$&ID \\ 
\hline
 &&&&&\\
1 & 1062.883 & 3158.63$\pm$0.20 & $\ge$3.02$^a$ & 1.9718 & L3-0 R(0)   \\
2 & 1063.460 & 3161.85$\pm$0.10 & $\ge$2.74$^a$ & 1.9732 & L3-0 R(1) \\
3 & 1064.606 & 3165.01$\pm$0.12 & $\ge$1.47$^a$ & 1.9729 & L3-0 P(1) \\
4 & 1064.995 & 3166.94$\pm$0.30 & 0.77$\pm$0.16 & 1.9737 & L 3-0 R(2) \\
5 & 1066.901 & 3171.78$\pm$0.11 & 0.79$\pm$0.15 & 1.9729 & L 3-0 P(2) \\
6 & 1067.478 & 3173.80$\pm$0.11 & 0.79$\pm$0.15 & 1.9732 & L 3-0 R(3) \\
7 & 1070.142 & 3182.18$\pm$0.10 & 0.74$\pm$0.12 & 1.9736 & L 3-0 P(3) \\
8 & 1070.898 & 3184.14$\pm$0.19 & 0.33$\pm$0.10 & 1.9733 & L 3-0 R(4) \\
9 & 1074.313 & 3194.04          & $\le$ 0.202$^b$& 1.9731 & L 3-0 P(4) \\
10& 1085.382 & 3226.95          & $\le$ 0.248$^b$& 1.9731 & L 3-0 P(6) \\
11& 1086.626 & 3230.65          & $\le$ 0.256$^b$& 1.9731 & L 3-0 R(7) \\
12& 1077.138 & 3202.32$\pm$0.10 & $\ge$1.99$^a$  & 1.9730 & L 2-0 R(0) \\
13& 1077.678 & 3204.27$\pm$0.10 & $\ge$2.02$^a$ & 1.9733 & L 2-0 R(1) \\
14& 1078.926+1079.226 & 3208.14$\pm$0.10 & 1.89$\pm$0.20 & 1.9731 & L 2-0 P(1)+R(2)\\
15& 1081.265 & 3214.68$\pm$0.07 & 0.87$\pm$0.10 & 1.9731 & L 2-0 P(2) \\
16& 1088.794 & 3237.09          & $\le$ 0.20$^b$ &1.9731& L 2-0 P(4) \\
17& 1100.016 & 3270.46          & $\le$ 0.22$^b$ & 1.9731 & L 2-0 P(6) \\
18& 1092.194 & 3247.02$\pm$0.14 & $\ge$2.38$^a$ & 1.9729 & L 1-0 R(0) \\
19& 1092.732 & 3249.27$\pm$0.08 & $\ge$2.05$^a$ & 1.9735 & L 1-0 R(1) \\
20& 1094.052+1094.244 & 3253.13$\pm$0.06 & 2.08$\pm$0.14&1.9731 & L 1-0 P(1)+R(2) \\ 
21& 1099.788 & 3269.66$\pm$0.08 & 0.66$\pm$0.09 & 1.9730 & L 1-0 P(3) \\
22& 1100.165 & 3270.90          & $\le$ 0.22$^b$& 1.9731 & L 1-0 R(4) \\
23& 1104.084 & 3282.24$\pm$0.18 & 0.18$\pm$0.07& 1.9728 & L 1-0 P(4) \\
24& 1104.547 & 3283.93          & $\le$ 0.16$^b$ & 1.9731 & L 1-0 R(5) \\
25& 1108.128+1108.634 & 3295.23 & $\ge$2.04$^a$  & 1.9731 & L 0-0 R(0)+R(1)\\
26& 1112.495+1112.584 & 3207.59 & 1.25$\pm$0.13  & 1.9731 & L 0-0 P(2)+R(3)\\
27& 1120.246+1120.399 & 3331.10 & 0.34$\pm$0.08 & 1.9731 & L  0-0 P(4)+R(5)\\
28& 1125.539+1125.725 & 3346.65 & $\le$ 0.16$^b$ & 1.9731 & L 0-0 P(5)+R(6)\\
&&&&&\\
\hline

\end{tabular}

\medskip
\noindent $^a$ Lower limits due to the  probable existence of  the damping wings
of these lines

\noindent $^b$ 2$\sigma$ upper limits


\newpage
\rm
\footnotesize
\begin{flushleft}
Table 2. H$_2$ Column Densities and Excitation Temperatures in the z = 1.9731 Damped Ly\al\ Absorber 
\normalsize
\end{flushleft}

\begin{tabular}{llll}  \hline\hline
Level & N(cm$^{-2}$)$^a$ &  T$_{ex}$(K) & J$_{ex}$ \\ 
\hline
 &&&\\
J = 0 &   3.8$(\pm 1.2)\times 10^{19}$ &         &    \\
J = 1 &   3.0$(\pm 1.0)\times 10^{19}$ & 70 $\pm$ 13     & 0-1\\
J = 2 &   3.8$^{+4.8}_{-1.1}\times 10^{17}$ & 82$^{+17}_{-6}$    & 0-2\\
J = 3 &   1.4$^{+2.1}_{-0.6}\times 10^{17}$ & 208$^{+168}_{-44}$    & 2-3\\
J = 4 &   2.5$^{+1.2}_{-0.8}\times 10^{15}$ & 209$^{+50}_{-16}$     & 2-4\\
J = 5 &   $\le 1.6\times 10^{15}$ & $\le$ 271    & 2-5\\
J = 6 &   $\le 1\times 10^{15}$  & $\le 430$ & 2-6\\
J = 7 &   $\le 9\times 10^{14}$ & $\le 497$ & 2-7\\
&&&\\
\hline

\end{tabular}

\noindent $^a$ 2$\sigma$ upper limits given for J = 5, 6, 7

\newpage
\tiny
\oddsidemargin=-0.7in
\begin{flushleft}
Table 3. Comparison between the z = 1.9731 Absorber and $\zeta$ Oph and
$\xi$ Per clouds
\end{flushleft}
\begin{tabular}{llllllllllll}
\hline
\hline
Name & lg N(H$_2$)&lg N(H I)&lg $f$ & [Zn/H] & [Fe/H] & T$_{01}$  &
T$_{02}$ & T$_{ex}$& n(H) & $\beta_0$& b\\ 
& (\cm) & (\cm)  & & & & (K) &(K) &(K) & (cm$^{-3}$)& ($10^{-10}$s$^{-1}$)&(\kms)\\
&&&&&&&&&&&\\
Q 0013  & 19.86 $\pm$ 0.10 & 20.70$\pm 0.05$ &$-$0.66 $\pm$ 0.10 & $-0.80\pm 
0.08^c$ & $-$1.95$\pm 0.01$& 70 $\pm$ 13 & 82$^{+17}_{-8}$ & 200-500 & $\sim$ 300 & $\sim$ 67& 15 $\pm$ 2\\
$\zeta$ Oph$^a$ & 20.65 & 20.72 & $-$0.20 & $>-$0.81 & $-$2.38 & 54 & 84 & 324 & 700 & 
$\sim$ 10$^d$ & 3.8\\
$\xi$ Per$^b$  & 20.53 & 21.11 & $-$0.46 & $-$0.65 & $-$2.24& 71& 78 & 384 &  300 & 50& 5  \\
&&&&&&&&&&&\\
\hline

\end{tabular}

\noindent $^a$ Results are from Spitzer {\em et al.} 1974; Savage  {\em et al.}   1977; Jura 
1975; Savage {\em et al.}   1992.

\noindent $^b$ Data are from Spitzer {\em et al.}   1974; Savage {\em et al.}   1977; Jura 1975; Cardelli {\em et al.}   1991.

\noindent $^c$ Based on the f-value from Bergeson {\em et al.} (1993) 

\noindent $^d$ Federman {\em et al.} (1995)

\newpage
\pagestyle{empty}
\textwidth 200mm
\textheight 280mm
\topmargin -10 mm
\oddsidemargin -10mm

\normalsize
\centerline{\bf Figure Captions}

\noindent Figure 1. -- (a). Spectrum of QSO  0013$-$004 obtained with the MMT Blue
 Channel
Spectrograph. The dotted line is the synthetic absorption line spectrum of  
the $\upsilon$ = 0 - 0, 1 - 0, 2 - 0 and 3 -  0  Lyman bands associated
with the z = 1.9731 damped  Ly\al\ absorber. Strong absorption lines 
No. 1, 2, 3 are
identified as N II \lam\ 1083.99 \AA\ at z = 1.9673, 1.9714, 1.9731, 
respectively, consistent with other metal line identifications (Ge {\em et al.}   1997).
 The strong
line No. 4 is identified as H I Ly\al\ line at z = 1.7135, which shows C IV
\lam\lam\ 1548,1550 lines in previous observed QSO spectra (Ge {\em et al.}   1997). (b). Cross correlation between the QSO spectrum and the 
synthetic H$_2$ absorption line spectrum shown  in Fig. 1. 
 The peak near  zero velocity has the largest
amplitude  in the testing range $-$16,000 to 16,000 \kms. The two second highest
peaks are evidently caused by shifting two spectra with one Lyman band 
separation ($\sim$  47  \AA\ at the  observed  frame).

\noindent Figure 2. --  The damped Ly$\alpha$ absorption line at z = 1.9731. 
The three curves are  profiles with $N(H I) = (4.5,5.0,5.5)\times 10^{20}$
 \cm\ and b = 50 \kms. 

\noindent Figure 3. -- Curves of growth for H$_2$ (0,0) Lyman band with b =
10,  13, 15, 17 and 20 \kms. The lines  from J = 4, 5 are on the  linear part, while
those  from J = 2, 3 are  on the flat part of the curve of growth. 
All lines are  consistent with b = 15$\pm 2$ \kms.

\newpage

\begin{figure}
\plotone{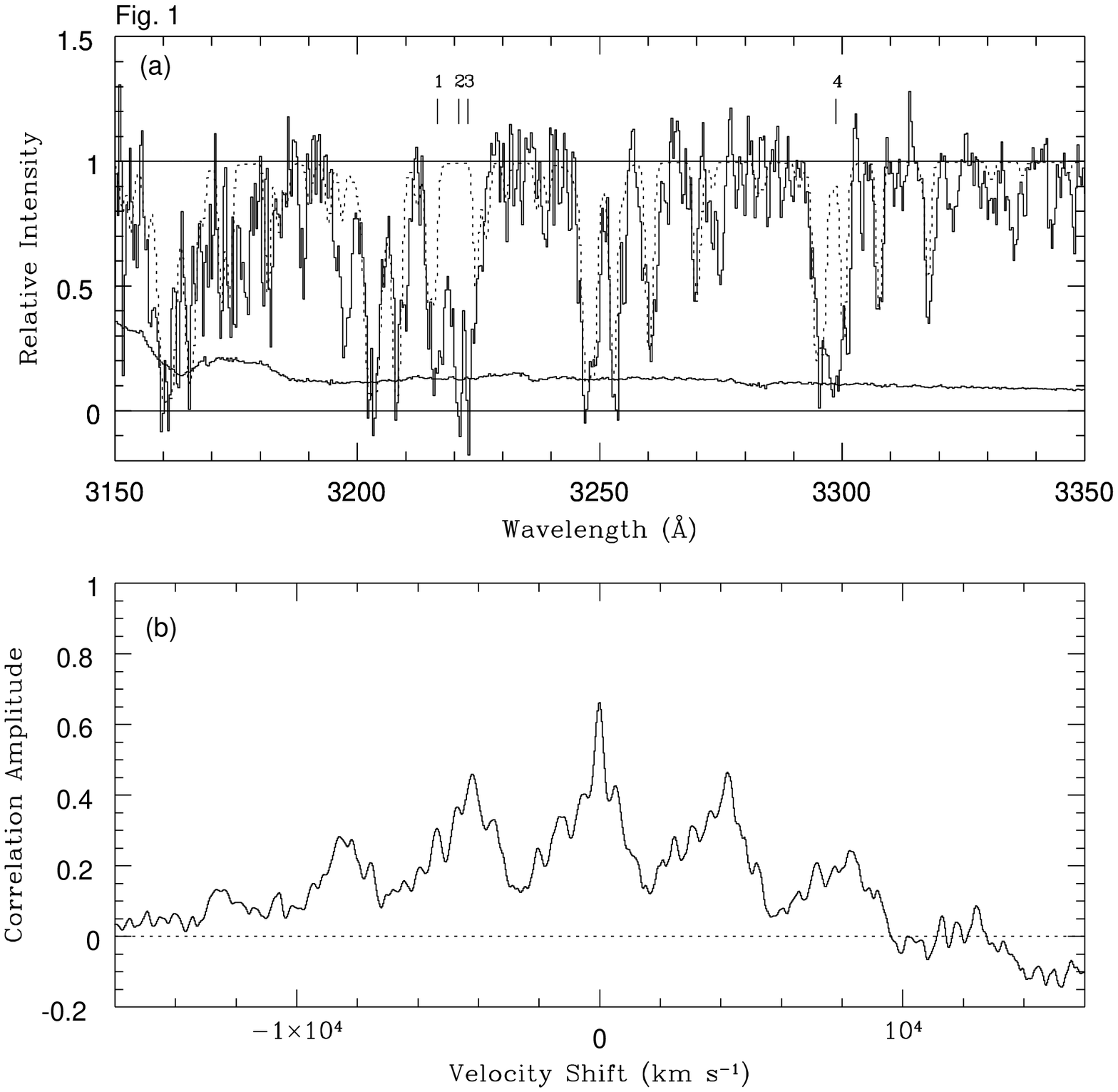}
\end{figure}

\newpage

\begin{figure}
\plotone{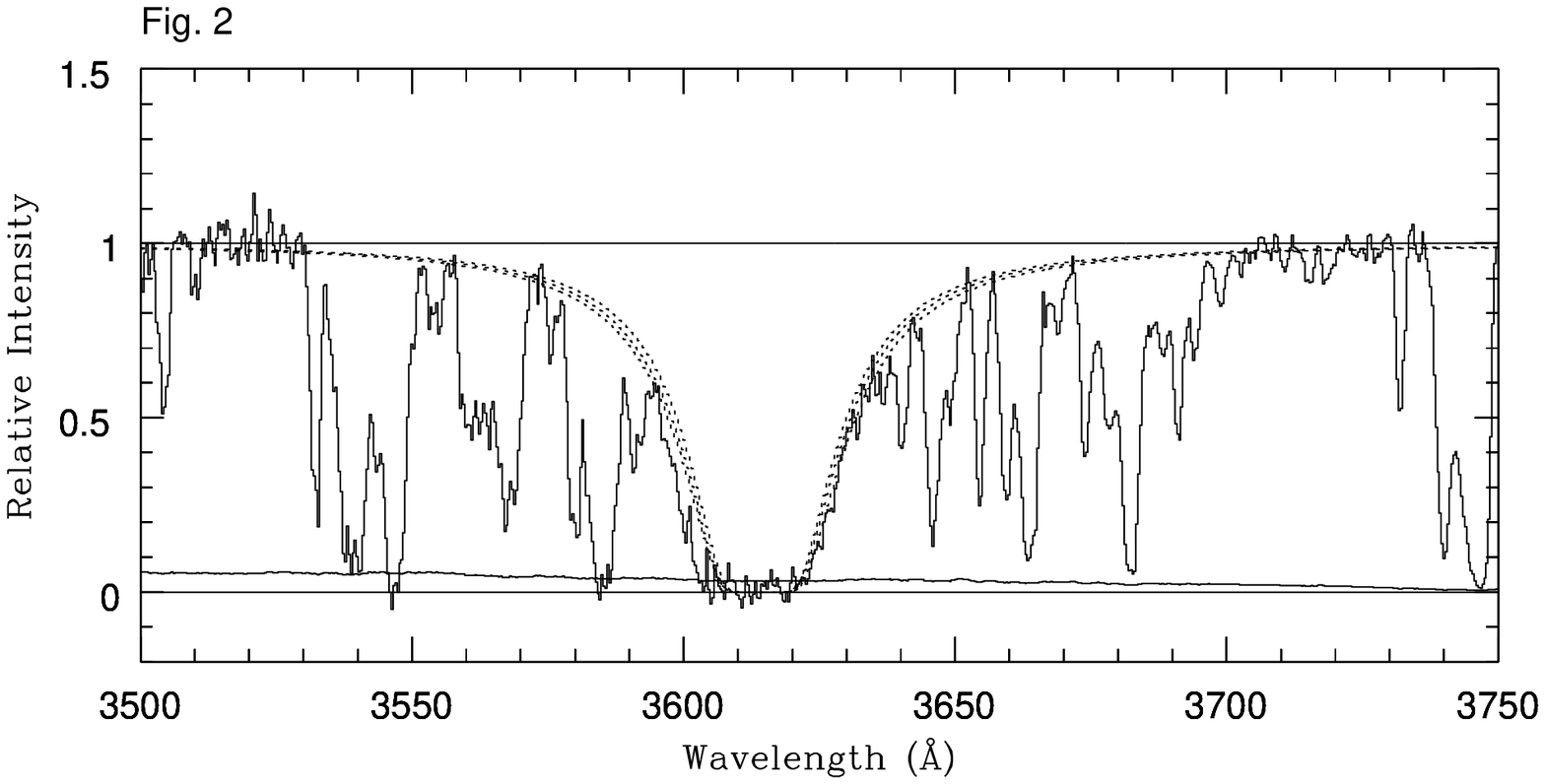}
\end{figure}

\newpage

\begin{figure}
\plotone{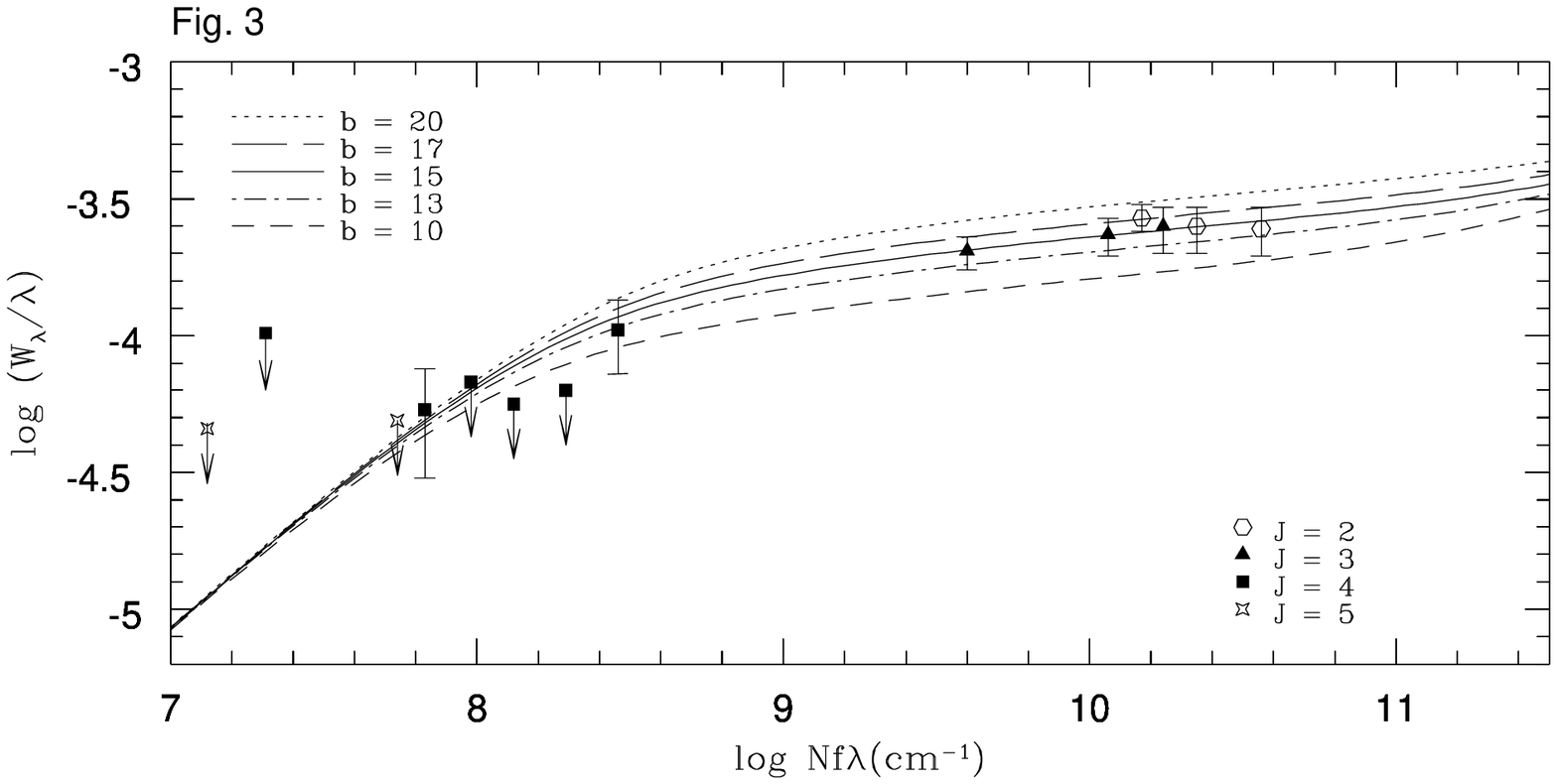}
\end{figure}

\end{document}